\newcommand{\kl}[3]{\mbox{$\rm #1$}^{\mu\nu , \alpha\beta}_{#2}(#3)}

\def\be{\begin{equation}}
\def\te{\end{equation}}
\def\bea{\begin{eqnarray}}

\def\tea{\end{eqnarray}}

\def\k{\kappa}

\def\m{\mu}
\def\n{\nu}

\newskip\humongous \humongous=0pt plus 1000pt minus 1000pt

\newif\ifdtup

\def\ha{{1\over 2}}

\documentstyle[11pt]{article}

\makeatletter                    
\@addtoreset{equation}{section}  
\makeatother                     

\begin{document}
\title{Black Hole Fluctuations and Backreaction in Stochastic Gravity}
\author{
Sukanya Sinha\\
{\small Indian Statistical Institute, Bangalore Centre,
 8th Mile, Mysore Road, Bangalore-560059, INDIA}\\
Alpan Raval\\
{\small Department of Mathematics, Claremont Graduate University,
121 E. Tenth St., Claremont, CA 91711,}\\ {\small Keck
Graduate Institute, 535 Watson Drive, Claremont, CA 91711.}\\
B. L. Hu\\
{\small Department of Physics, University of Maryland, College
Park, MD 20742}}
\date{({\small umdpp 03-20, Oct 4, 2002})}
\maketitle \centerline {\small - To appear in a special issue of
Foundations of Physics  {\it Thirty Years of Black Hole Physics}
edited by L. Horwitz.}
\begin{abstract}
\noindent We present a framework for analyzing black hole
backreaction from the point of view of quantum open systems using
influence functional formalism. We focus on the model of a black
hole described by a radially perturbed quasi-static metric and
Hawking radiation by a conformally coupled massless quantum
scalar field. It is shown that the closed-time-path (CTP)
effective action yields a non-local dissipation term as well as a
stochastic noise term in the equation of motion, the
Einstein-Langevin equation.
Once the thermal Green's function in a Schwarzschild background
becomes available to the required accuracy the strategy described
here can be applied to obtain concrete results on backreaction.
We also present an alternative derivation of the CTP effective
action in terms of the Bogolyubov coefficients, thus making a
connection with the interpretation of the noise term as measuring
the difference in particle production in alternative histories.
\end{abstract}
\noindent {\it This essay is dedicated to my friend and colleague,
Professor Jacob Bekenstein, with great admiration and fond
memories since our graduate student days. Jacob's pioneering
contributions helped establish a new discipline, black hole
thermodynamics, and ushered in a new era of research in
gravitation, statistical and quantum physics which probes the
nature of spacetime and matter at the most fundamental level.} --
B. L. Hu

\newpage
\section{Introduction}
\label{sec:intro}

Quantum field theory in curved spacetime \cite{qftcst} is a
subject explored from the mid 60's to the 70's which deals with
important effects such as cosmological particle creation
\cite{cpc}  and Hawking radiation from black holes \cite{bhpc}.
The effect of particle creation on the background spacetime --
known as the backreaction problem -- becomes important when the
energy reaches the Planck scale, as in the early universe
\cite{cpcbkr} and at the final stages of black hole evolution
\cite{HI,Bardeen,York,HK,HKY,AHWY,JMO,MPP,AHS,AHL,HLA}.
Investigation of backreaction problems started with the study of
the renormalization or regularization of the energy momentum
tensor in curved spacetimes in the late 70's. Because of the
higher symmetry in cosmological spacetimes, backreaction studies
of processes therein have progressed further than the
corresponding black hole problems, which to a large degree is
still concerned with finding the right approximations for the
energy momentum tensor \footnote{The latest important work is
that of Hiscock, Larson and Anderson \cite{HLA} on backreaction
in the interior of a black hole, where one can find a concise
summary of earlier work. To name a few of the important landmarks
in this endeavor (this is adopted from \cite{HLA}), Howard and
Candelas \cite{HowCan,How} have computed the stress-energy of a
conformally invariant scalar field in the Schwarzschild geometry.
Jensen and Ottewill \cite{JenOtt} have computed the vacuum
stress-energy of a massless vector field in Schwarzschild.
Approximation methods have been developed by Page, Brown, and
Ottewill \cite{Page82,BroOtt,PBO} for conformally invariant
fields in Schwarzschild spacetime, Frolov and Zel'nikov
\cite{FZ1} for conformally invariant fields in a general static
spacetime, Anderson, Hiscock and Samuel \cite{AHS} for massless
arbitrarily coupled scalar fields in a general static spherically
symmetric spacetime. Furthermore the DeWitt-Schwinger
approximation has been derived by Frolov and
Zel'nikov\cite{FZ82,FZ84} for massive fields in Kerr spacetime,
Anderson Hiscock and Samuel \cite{AHS} for a general (arbitrary
curvature coupling and mass) scalar field in a general static
spherically symmetric spacetime and have applied their method to
the Reissner-Nordstr\"{o}m geometry \cite{AHL}.} for even the
simplest spacetimes such as the spherically symmetric family
including the important Schwarzschild metric. Though arduous and
demanding, the effort continues on because of the importance of
backreaction effects of Hawking radiation in black holes. They
are expected to address some of the most basic issues such as
black hole thermodynamics \cite{BekHaw,smbhent,strbhent} and the
black hole end-state and information loss puzzles
\cite{Pagereview}.

The most significant developments in the implementation and
physical aspects of backreaction problems in the 80's as
exemplified first in cosmological spacetimes are perhaps the
introduction of an effective action which yields real and causal
equations of motion. (For advances in the axiomatic theoretical
aspects, see the work of, e.g., Kay and Wald, Flanagan and Wald
\cite{KayWal,FlaWal}.)  This is known as the closed-time-path
(CTP) or the Schwinger-Keldysh method \cite{ctp}. From this one
can identify dissipative effects in an unambiguous manner and in
the true statistical mechanical sense. In the 90's, the most
significant development was perhaps the introduction of quantum
open systems concepts \cite{qos} and the influence functional
method \cite{if}, which enables one to identify the origin of
quantum noise and recognize the importance of fluctuations. The
latter effort has crystallized into a new theory known as
stochastic semiclassical gravity (SSG) which is the next stage
beyond semiclassical gravity (SCG) towards quantum gravity. These
developments were summarized in reviews, (e.g.,
\cite{Banff,stogra,HVErice,CQGrev,LivRev}) to which the reader is
referred for details and further references.

Here we wish to address the old black hole backreaction problems
with these newer insights. It is not our intention to seek better
approximations for the regularized energy momentum tensor, but to
point out new ingredients lacking in the existing framework based
on SCG. In particular one needs to consider both the dissipation
and the fluctuations aspects in the back reaction effects of
particle creation or vacuum polarization.

In a short note \cite{Vishu} we discussed the formulation of the
problem in this new light, commented on the shortcomings of
existing works, and sketched the strategy behind our own approach
to the black hole fluctuations and backreaction problem. Here we
continue the train of thought with more details for the class of
quasi-static black holes, leaving the more demanding dynamical
collapse problem to a later exposition. Thus we only address the
first set of major issues mentioned above.

 From the new perspective provided by statistical field theory and
stochastic gravity, it is not difficult to project that
backreaction effect is the manifestation of a fluctuation-
dissipation relation (FDR) \cite{FDR}. This was first conjectured
by Candelas and Sciama \cite{CanSci} for a dynamic Kerr black hole
emitting Hawking radiation.  Mottola \cite{Mottola} derived  from
linear response theory a FDR for a static black hole (in a box) in
quasi-equilibrium with its radiation. In \cite{Vishu} we showed
the shortcomings of these two important earlier work. In short,
Mottola's linear response FDR is based on the assumption of a
specified background spacetime (static in this case) and state
(thermal) of the matter field(s). But linear response theory is
not suitable for backreaction investigations because the spacetime
and the state of matter should be determined in a self-consistent
manner by their dynamics and mutual influence.

 For Candelas and Sciama \cite{CanSci}, the classical formula they
showed relating the dissipation in area linearly to the squared
absolute value of the shear amplitude is suggestive of  a
fluctuation-dissipation relation. When the gravitational
perturbations are quantized (they choose the quantum state to be
the Unruh vacuum) they argue that it approximates a flux of
radiation from the hole at large radii. Thus the dissipation in
area due to the Hawking flux of gravitational radiation is
allegedly related to the quantum fluctuations of gravitons.
Our criticism is that their's is not a FDR in the truly
statistical mechanical sense because it does not relate
dissipation of a certain quantity (in this case, horizon area) to
the fluctuations of {\it the same quantity}. To do so would
require one to compute the two point function of the area, which,
being a four-point function of the graviton field, is related to
a two-point function of the stress tensor. The stress tensor is
the true ``generalized force'' acting on the spacetime via the
equations of motion, and the dissipation in the metric must
eventually be related to the fluctuations of this generalized
force for the relation to qualify as a FDR.

From this reasoning, we see that the stress energy bi-tensor and
its vacuum expectation value known as the noise kernel, are the
new ingredients in backreaction considerations. But these are
exactly the centerpiece in stochastic gravity. Therefore the
correct framework to address semiclassical backreaction problems
is stochastic gravity theory, where fluctuations and dissipation
are the equally essential components. The calculation of 4-point
functions of the metric perturbations $h_{ab}$ in Minkowski
spacetime has been carried out by Martin, Roura and Verdaguer
\cite{MarVer,RouVer}, for thermal fields in black hole spacetime
and scalar fields in general spacetimes by Campos, Hu and
Phillips \cite{CamHu,PH1,PH2,PH3}. Returning to the point of
understanding backreaction as manifestation of FDR, a generalized
FDR expression relating dissipation (of anisotropy in the case
under their study) and fluctuations (measured by particle numbers
created in neighboring histories) was obtained by Hu and Sinha
earlier \cite{HuSin} for cosmological backreaction problems.

In this note we present a sketch of the stochastic gravity theory
and quantum open systems as  applied to black hole backreaction
problems focusing only on the quasi-static class. Campos and Hu
\cite{CamHu} treated the far field case, while here we consider
the near horizon case. We use the same approach and show the
appearance of two new terms in the stochastic effective action
and equation of motion for backreaction: nonlocal dissipation and
(generally colored) noise kernels. When the noise average of this
Einstein-Langevin equation is taken,  York's \cite{York}
semiclassical equations for radially perturbed quasi-static black
holes is recovered. We can only present the overall structure of
the theory and the strategy of our approach, but not the details,
because the Green function for a scalar field in the
Schwarzschild metric comes only in an approximate form (e.g. Page
approximation \cite{Page82}, which, though reasonably accurate
for the stress tensor, fails poorly for the noise kernel
\cite{PH2}). In addition we present an alternative way to obtain
the CTP effective action, i.e., by expressing it in terms of the
Bogolyubov coefficients, which measure not only the number of
particles created, but also the difference of particle creation
in alternative histories. We see this as a useful avenue to
explore the noise and fluctuations issues in black hole physics.

We begin in Sec. 2 with a summary description of quantum open
systems followed by a sketch of stochastic gravity theory. In
Section 3 we use a simple case of a scalar field in a static,
spherically perturbed Schwarzschild spacetime (York's model) to
illustrate how to calculate the stochastic effective action and
derive the Einstein-Langevin equation. In Sec. 4 we derive an
expression for the CTP effective action in terms of the Bogolyubov
coefficients. This is an original result not yet published
anywhere. We conclude in Sec. 5 with some discussions of the
potentials and limitations of our work, and how it relates to
other work on black hole fluctuations.


\section{Stochastic Approach to Backreaction Problems}

\subsection{Quantum Open Systems}
\label{sec:open} Here we give a brief schematic summary of the
Influence Functional method of analyzing quantum open systems to
illustrate the concept, the details can be found in {\cite{if}}.
This approach is designed to deal with the situation in which the
system $S$ described , say, by the degrees of freedom $x$ is
interacting with an environment $E$, described by the degrees of
freedom $q$. \footnote{We are labeling the degrees of freedom of
the system and the environment by single letters $x$ and $q$ with
the understanding that they can represent multiple or even
infinite degrees of freedom, e.g. corresponding to a field.} The
full closed quantum system $ S + E$ is described by a density
matrix $\rho (x,q;x',q',t)$. If we are interested only in the
state of the system as influenced by the overall effect, but not
the precise state of the environment, i.e, the dynamics of the
open system, then the reduced density matrix
$\rho_r(x,x',t)=\int~dq~\rho (x,q;x',q,t)$ would provide the
relevant information. (The subscript $r$ stands for reduced.)
Assuming that the action of the coupled system decomposes as
$S=S_s[x]+S_e[q]+S_{int}[x,q]$, and that the initial density
matrix factorizes (i.e., takes the tensor product form), $\rho
(x,q;x',q',t_i)=\rho_s(x,x',t_i)\rho_e(q,q',t_i)$,  the reduced
density matrix is given by \be \rho_r(x,x',t)
=\int\limits_{-\infty}^{+\infty}dx_i\int\limits_{-\infty}^{+\infty}dx'_i~
 \int_{x_i}^{x_f} Dx~ \int_{x'_i}^{x'_f} Dx '~
e^{i(S_s[x]-S_s[x']+S_{IF}[x,x',t])} ~\rho_r(x_i,x'_i,t_i~)
\label{pathint}
\te
where $S_{IF}$  is the
influence action related to the influence functional
$\cal F$ defined by

\be
{\cal F}[x, x'] \equiv e^{iS_{IF}[x,x',t]}\equiv\int~dq_f~dq_i~ dq_i'~
  \int_{q_i}^{q_f}Dq~\int_{q'_i}^{q_f}Dq'~
e^{i(S_e[q]+S_{int}[x ,q]-S_e[q']-S_{int}[x ',q'])}
\rho_e(q_i,q'_i,t_i).
\label{SIF}
\te
$S_{IF}$ in general is complex. Retaining only
quadratic terms (an approximation which covers many of the
interesting applications that we will consider later), we may write

\be S_{IF}(x,x')=\int~dt~dt'~\{ {1\over
2}(x-x')(t)D(t,t')(x+x')(t') +{i\over
2}(x-x')(t)N(t,t')(x-x')(t')\} \label{SIFquad} \te where $D$ and
$N$ stand for the real dissipation and noise kernels respectively.
Note that in this quadratic order approximation, the influence
action $ S_{IF}(x,x')$ is related to the closed-time-path(CTP) or
in-in effective action (for details on the CTP effective action
see \cite{ctp})  $\Gamma_{CTP}[x,x']$ through \cite{ctpif} \be
\Gamma_{CTP}[x,x'] = S[x] - S[x'] + S_{IF}[x,x'].
\label{CTPIFconn} \te The equation of motion obtained from the
CTP effective action for the expectation values are clearly seen
to be real and causal \cite{ctp}. They read \be \left.{\delta\over
\delta x(t)}\Gamma_{CTP}[x, x']\right|_{x' = x = {\bar x}} = 0
\label{CTPem} \te From the influence functional a Langevin
equation for the system dynamics may be derived by a formal
procedure, first introduced by Feynman and Vernon \cite{if}, which
consists of introducing  a Gaussian stochastic source $\xi(t)$
with $<\xi(t)>_{\xi} = 0$ and $<\xi(t)\xi(t')> = N(t,t')$ and
defining an improved or stochastic effective action as \be
S_{eff}[x,x';\xi] = S_s[x] - S_s[x'] + {\cal{R}}S_{IF}[x,x'] +
\xi . (x-x'). \label{Seffxi} \te such that $\left <
e^{iS_{eff}[x,x';\xi]}\right >_{\xi} = e^{i\Gamma_{CTP}[x,x']}$.
This leads to equations of motion with a stochastic force: \be
\left.{\delta S_{eff}[x,x';\xi]\over \delta x}\right|_{x =x'} = 0.
\label{stochem} \te or, equivalently, \be \left.{\delta
\Gamma_{CTP}[x,x';\xi]\over \delta x}\right|_{x =x'} = 0.
\label{CTPstoch} \te
The equation of motion obtained from (\ref{stochem}) using
(\ref{Seffxi}) is  \be {\partial S_s\over\partial
x(t)}+\int~dt'~\gamma (t,t') {d\ x(t')\over dt'}= \xi
\label{langevin} \te where $D(t,t')=-\partial_{t'}\gamma (t,t')$.
Being now in the form of a Langevin equation ,  the physical
meaning of the $\gamma$ and $N$ kernels in Eq. (\ref{langevin})
become clearer. Both the terms involving $\gamma$ and $\xi$
represent the backreaction of the environment on the system.
However, $\gamma$ (or more properly the odd part of $\gamma$) is
associated with dissipation and $\xi$ is a stochastic noise term
associated with random fluctuations of the system exactly as the
terms are interpreted in the context of Brownian motion.
Averaging (\ref{langevin}) over the noise using the appropriate
probability distribution will give the semiclassical equation of
motion for the mean value of $x$. The noise and dissipation
originating from a closed system (as is done here, as opposed to
being put in by hand) are in general related by a set of
generalized fluctuation-dissipation relations, (FDR) which can be
represented by a linear, non-local relation of the form, \be
N(t-t') = ~\int~d(s -s')K(t-t',s-s')\gamma(s -s') \label{FDR} \te
We will discuss various aspects of the FDR in greater detail in
later sections. To keep the discussion simple, we have written the
noise and dissipation kernels in terms of single scalar
functions. However, the method is general enough to encompass
multiple noise and dissipation kernels and cases where the
kernels are tensorial, as in our later examples.

\subsection{Stochastic Semiclassical Gravity}
\label{sec:stogra}

The framework described in the previous section is general enough
to encompass any situation where we want to describe the
semiclassical dynamics of the ``system" degrees of freedom with
the backreaction of the ``environment" degrees of freedom
incorporated self-consistently. One has to make sure there is a
discrepancy parameter in the problem to define our interested
subsystem and justify separating it from its environment.
Coarse-graining the environment and incorporating its effect on
the subsystem turns it into an open system. In the context of
semiclassical gravity our ``system" of interest is a spacetime
metric coupled to the ``environment" of a quantum scalar field
(the discrepancy parameter being the Planck mass). The dynamic
classical spacetime metric creates particles of the quantum field
and these in turn provide a backreaction on the space time metric
to alter its dynamics in response. This is captured in the so
called semiclassical Einstein equations (SEE) which take the form
 \be \label{semi} G_{\m\n} (g_{\alpha \beta})
= \k \langle T_{\m\n} \rangle_q \te where $T_{\m\n} $ is the
energy momentum tensor of, say, a free scalar field $\Phi$,
$G_{\mu\nu}$ is the Einstein tensor, $\kappa = 8\pi G_N$ , $G_N$
being the Newton's constant. Here $\langle\,\rangle_q$ denotes
expectation value taken with respect to some quantum state with
symmetry commensurate with that of the background spacetime.
Studies of semiclassical Einstein equation have been carried out
in the last two decades  by many authors for cosmological
\cite{cpcbkr} and black hole spacetimes
\cite{HI,Bardeen,York,HK,HKY,AHWY,JMO,MPP,AHS,AHL}. In the
analogy with the open system dynamics described in Section
\ref{sec:open}: Eq.(\ref{semi}) is equivalent to Eq.(\ref{CTPem})
where the degrees of freedom $x$ of the system are identified
with the metric $g_{\alpha\beta}$ and those of the environment
$q$ are identified with the scalar field $\Phi(x)$. However, from
the discussion in the last section it is also clear that Eq.
(\ref{CTPem}) , and hence also the semiclassical Einstein Eq.
(\ref{semi}) results on averaging the full Langevin-type Eq.
(\ref{stochem}) over noise. Thus the semiclassical Einstein
equation incorporates the dissipation but misses out the
fluctuation aspect of the backreaction. The recognition of this
crucial point \cite{Physica} ushered in a new theory known as
stochastic semiclassical gravity (SSG) , (or stochastic gravity in
short as there is no confusion in this context as to where the
stochasticity originates). Aided by the concept of open systems
and the techniques of influence functional and the CTP effective
action, stochastic gravity  is the new framework where one should
consider backreaction problem. We shall do this for the black
hole problem here. Stochastic gravity can treat noise and
fluctuations from particle creation on the same footing as
dissipation. Spacetime dynamics is now governed by a stochastic
generalization of the semiclassical Einstein equation known as
the Einstein-Langevin equation,  the analog of Eq.
(\ref{langevin}) in the context of semiclassical gravity. The
conventional theory of semiclassical gravity (SCG) with sources
given by the vacuum expectation value of the energy momentum
tensor is viewed as a mean field approximation of this new
theory. Schematically the Einstein-Langevin equation takes on the
form
\begin{eqnarray}
   \tilde G_{\mu\nu}(x)
        &=& \k  \left( T_{\mu\nu}^{\rm c} +  T_{\mu\nu}^{\rm qs}
\right),
           \nonumber \\
   T_{\mu\nu}^{\rm qs}
        &\equiv& \langle T_{\mu\nu} \rangle_{\rm q} +  T_{\mu\nu}^{\rm
s}
\label{eq:effective stress tensor}
\end{eqnarray}
Here, $T_{\mu\nu}^c$ is due to classical matter or fields,
$\langle T_{\mu\nu} \rangle_q$ is the vacuum expectation value of
the stress tensor of the quantum field, and  $T_{\mu\nu}^{\rm qs}$
is a new stochastic term which is related to the fluctuations of
$T_{\mu\nu}$ in the vacuum state. Taking the average of
(\ref{eq:effective stress tensor}) with respect to the noise
distribution will lead to the conventional semiclassical Einstein
equation. It is clear in this context why SCG is regarded as a
mean field theory. For a detailed derivation and solution of the
Einstein-Langevin equation in the context of semiclassical
cosmology the reader is referred to \cite{CamVer}.

\section{Black Holes: Backreaction and Fluctuations}

Let us now come to the focus of this article, namely the issue of
how to incorporate the new stochastic features of semiclassical
gravity in the context of black holes. Our interest is to employ
the full force of the formalism developed in section
\ref{sec:open}  to this problem and see the effect of the
stochastic noise term on black hole backreaction. In this article
we will sketch the strategy of this program rather than describe
detailed calculations which are still in progress. We focus on
the simpler class of problems of a quasi-static black hole in
quasi-equilibrium (a box is required) with its Hawking radiation
described by a scalar field. The goal is to obtain an influence
action analogous to (\ref{Seffxi}) for this model of a black hole
coupled to a scalar field and to be able to derive an
Einstein-Langevin equation analogous to (\ref{langevin}) from it.
To this end we will first proceed to derive the CTP effective
action $\Gamma_{CTP}$ for this case and then use it via
(\ref{CTPIFconn}) to derive the stochastic influence action
({\ref{Seffxi}).

Let us make a modest beginning by considering the simplest model
of this class described by a perturbed Schwarzschild metric. This
has been previously used by York \cite{York} to analyze black
hole backreaction. We do not intend to offer a \textit{better}
solution than what was done before, but to bring out the
\textit{new} physics unbeknown to researchers of backreaction
problems in SCG, in particular, noise and fluctuations and their
consequences.

\subsection{The Model}

In this model the black hole spacetime is described by a
spherically symmetric static metric with line element of the
following general form written in advanced time
Eddington-Finkelstein coordinates \be ds^2 =
g_{\mu\nu}dx^{\mu}dx^{\nu} = -e^{2\psi}\left(1 - {2m\over
r}\right)dv^2 + 2 e^{2\psi}dvdr + r^2~d{\Omega}^2
\label{ssmetric} \te where $\psi = \psi(r)$ and $m = m(r)$ , $ v
= t + r + 2Mln\left({r\over 2M} -1 \right)$ and $d{\Omega}^2$ is
the line element on the two sphere. Hawking radiation is
described by a massless, conformally coupled quantum scalar field
$\phi$ with the classical action \be S_m[\phi, g_{\mu\nu}] =
-{\ha}\int d^n x \sqrt{-g}[g^{\mu\nu}\partial_{\mu}\phi
\partial_{\nu}\phi + \xi(n) R{\phi}^2] \label{phiact} \te where
$\xi(n) = {(n-2)\over 4(n-1)}$ ($n$ is the dimension of
spacetime) and $R$ is the curvature scalar of the spacetime it
lives in.

Let us consider linear perturbations $h_{\mu\nu}$ off a
background Schwarzschild metric metric $g^{(0)}_{\mu\nu}$ \be
g_{\mu\nu} = g^{(0)}_{\mu\nu} + h_{\mu\nu} \label{linearize} \te
with standard line element \be (ds^2)^0 = \left( 1 - {2M\over
r}\right)dv^2 + 2dvdr + r^2d{\Omega}^2 \label{schwarz} \te We
look for this class of perturbed metrics in the form given by
(\ref{ssmetric}), (thus restricting our consideration only to
spherically symmetric perturbations): \be e^\psi \simeq  1+
\epsilon \rho(r) \label{rho} \te and \be m \simeq M[ 1 + \epsilon
\mu (r)] \label{mu} \te where ${\epsilon\over \lambda M^2} =
{1\over 3}a T_H^4 ;$ $ a ={{\pi}^2\over 30} ; \lambda =
90(8^4)\pi^2$. $T_H$ is the Hawking temperature. This particular
parametrization of the perturbation is chosen following York's
\cite{York} notation. Thus the only non-zero components of
$h_{\mu\nu}$ are \be h_{vv} = -\left((1 - {2M\over r})2\epsilon
\rho(r) + {2M\epsilon \mu (r)\over r}\right) \label{hvv} \te and
\be h_{vr} = \epsilon\rho (r) \label{hvr} \te So this represents
a metric with small static and radial perturbations about a
Schwarzschild black hole. The initial quantum state of the scalar
field is taken to be the Hartle Hawking vacuum, which is
essentially a thermal state at the Hawking temperature and it
represents a black hole in (unstable) thermal equilibrium with
its own Hawking radiation.

Before we go on to describe our strategy for treating
backreaction in this model, let us briefly recall York's analysis
so that it will facilitate later comparisons with our program.
York \cite{York} \footnote{See also work by Hochberg and Kephart
\cite{HK} for a massless vector field, Hochberg, Kephart and York
\cite{HKY} for a massless spinor field, and Anderson, Hiscock,
Whitesell, and York \cite{AHWY} for a quantized massless scalar
field with arbitrary coupling to spacetime curvature} essentially
considers the semiclassical Einstein equation \be G_{\m\n}
(g_{\alpha \beta}) = \k \langle T_{\m\n}\rangle \te with
$G_{\mu\nu} \simeq G^{(0)}_{\mu\nu} + \delta G_{\mu\nu}$ where
$G^{(0)}_{\mu\nu}$ is the Einstein tensor for the background
spacetime. The zeroth order solution gives a background metric in
empty space, i.e, the Schwarzschild metric. $\delta G_{\mu\nu}$
is the linear correction to the Einstein tensor in the perturbed
metric. The semiclassical Einstein equation  in this
approximation therefore reduces to \be \delta G_{\mu\nu}(g^{(0)},
h) = \kappa <T_{\mu\nu}> \label{pertscee} \te He proceeds to
solve this equation to first order by using the expectation value
of the energy momentum tensor for a conformally coupled scalar
field in the Hartle-Hawking vacuum in the unperturbed
(Schwarzschild) spacetime on the right hand side and $\delta
G_{\mu\nu}$ on the left hand side is calculated using (\ref{hvv})
and (\ref{hvr}). Unfortunately, no exact analytical expression is
available for the $<T_{\mu\nu}>$ in a Schwarzschild metric with
the quantum field in the Hartle-Hawking vacuum that goes on the
right hand side. York therefore uses the approximate expression
given by Page \cite{Page82} which is known to give excellent
agreement with numerical results. Page's approximate expression
for $<T_{\mu\nu}>$ was constructed using a thermal Feynman Green's
function obtained  by a conformal transformation of a WKB
approximated Green's function for an optical Schwarzschild metric.
York then solves the semiclassical Einstein equation
(\ref{pertscee}) self consistently to obtain the corrections to
the background metric induced by the backreaction encoded in the
functions $\mu(r)$ and $\rho(r)$. There was no mention of
fluctuations or its effects. As we shall see, in the language of
the previous section, the semiclassical gravity procedure which
York followed working at the equation of motion level, is
equivalent to looking at the noise-averaged backreaction effects.

\subsection{CTP Effective Action for the Black Hole}

In this section, we set up the CTP effective action for the model
described in the previous section. The treatment given in this
section closely follows reference  \cite{CamHu}. Using the metric
(\ref{schwarz}) (and neglecting the surface terms that appear in
an integration by parts) we have  the action for the scalar
field  written perturbatively as
\begin{equation}
   S_m[\phi,h_{\mu\nu}]
        \ = \  {1\over 2}\int d^nx{\sqrt{-g^{(0)}}}\ \phi
               \left[ \Box + V^{(1)} + V^{(2)} + \cdots
              \right] \phi,
\label{phipert}
\end{equation}
where the first and second order perturbative operators $V^{(1)}$ and
$V^{(2)}$ are given by
\begin{eqnarray}
V^{(1)}   & \ \equiv \ & - {1\over \sqrt{-g^{(0)}}}
\left\{ [\partial_\mu\left(\sqrt{-g^{(0)}}\bar h^{\mu\nu}(x)\right)]
                                \partial_\nu
                              +\bar h^{\mu\nu}(x)\partial_\mu
                              \partial_\nu
                            +\xi(n) R^{(1)}(x)
                     \right\},
               \nonumber \\
V^{(2)}
    & \ \equiv \ & - {1\over \sqrt{-g^{(0)}}}
\left\{ [\partial_\mu \left(\sqrt{-g^{(0)}}
\hat h^{\mu\nu}(x)\right)]
                              \partial_\nu
                            +\hat h^{\mu\nu}(x)\partial_\mu
                            \partial_\nu
                          -\xi(n) ( R^{(2)}(x)
                               +{1\over 2}h(x)R^{(1)}(x))\right\}.
\end{eqnarray}
In the above expressions, $R^{(k)}$ is the $k$-order term in the
pertubation $h_{\mu\nu}(x)$ of the scalar curvature $R$ and $\bar
h_{\mu\nu}$ and $\hat h_{\mu\nu}$ denote a linear and a quadratic
combination of the perturbation, respectively,
\begin{eqnarray}
   \bar h_{\mu\nu}
        &  \equiv  & h_{\mu\nu} - {1\over 2} h g^{(0)}_{\mu\nu},
                     \nonumber \\
   \hat h_{\mu\nu}
        &  \equiv  & h^{\,\, \alpha}_\mu h_{\alpha\nu}
                      -{1\over 2} h h_{\mu\nu}
                      +{1\over 8} h^2 g^{(0)}_{\mu\nu}
                      -{1\over 4} h_{\alpha\beta}h^{\alpha\beta}
g^{(0)}_{\mu\nu}.
   \label{eq:def bar h}
\end{eqnarray}
 From quantum field theory in curved spacetime considerations we
take the following action for the gravitational field (see
\cite{CamHu,MarVer} for more details)
\begin{eqnarray}
   S^{(div)}_g[g_{\mu\nu}]
        & \ = \ & {1\over\ell^{n-2}_P}\int d^nx\ \sqrt{-g}R(x)
                \nonumber \\
        &       & +{\alpha\bar\mu^{n-4}\over4(n-4)}
                   \int d^nx\ \sqrt{-g}
                   \left[ 3R_{\mu\nu\alpha\beta}(x)
                           R^{\mu\nu\alpha\beta}(x)
                         -\left( 1-360(\xi(n) - {1\over6})^2
                          \right)R(x)R(x)
                   \right].
\end{eqnarray}
The first term is the classical Einstein-Hilbert action and the
second term is the counterterm in four dimensions used  to
renormalize the divergent effective action. In this action
$\ell^2_P = 16\pi G$, $\alpha = (2880\pi^2)^{-1}$ and $\bar\mu$
is an arbitrary mass scale.

We are interested in computing the CTP effective action for the
model given by the form (\ref{phipert}) for the matter action and
when the field $\Phi$ is initially in the Hartle- Hawking vacuum,.
This is equivalent to saying that the initial state of the field
is described by a thermal density matrix at a finite temperature
$T = T_H$. The CTP effective action at finite temperature $T
\equiv 1/\beta$ for this model is given by (for details see
\cite{CamHu})
\begin{equation}
   \Gamma^\beta_{CTP}[h^\pm_{\mu\nu}]
        \ = \ S^{(div)}_g[h^+_{\mu\nu}]
             -S^{(div)}_g[h^-_{\mu\nu}]
             -{i\over2}Tr\{ \ln\bar G^\beta_{ab}[h^\pm_{\mu\nu}]\},
   \label{eq:eff act two fields}
\end{equation}
where $\pm$ denote the forward and backward time path of the CTP
formalism and $\bar G^\beta_{ab}[h^\pm_{\mu\nu}]$ is the complete
$2\times 2$ matrix propagator ($a$ and $b$ take $\pm$ values:
$G_{++},G_{+-}$ and $G_{--}$ correspond to the Feynman, Wightman
and Schwinger Greens functions respectively) with thermal boundary
conditions for the differential operator $\sqrt{-g^{(0)}}(\Box +
V^{(1)} + V^{(2)} + \cdots)$. The actual form of $\bar
G^\beta_{ab}$ cannot be explicitly given. However, it is easy to
obtain a perturbative expansion in terms of $V^{(k)}_{ab}$, the
$k$-order matrix version of the complete differential operator
defined by $V^{(k)}_{\pm\pm} \equiv \pm V^{(k)}_{\pm}$ and
$V^{(k)}_{\pm\mp} \equiv 0$, and $G^\beta_{ab}$, the thermal
matrix propagator for a massless scalar field in Schwarzschild
spacetime . To second order $\bar G^\beta_{ab}$ reads,
\begin{eqnarray}
   \bar G^\beta_{ab}
        \ = \  G^\beta_{ab}
              -G^\beta_{ac}V^{(1)}_{cd}G^\beta_{db}
              -G^\beta_{ac}V^{(2)}_{cd}G^\beta_{db}
              +G^\beta_{ac}V^{(1)}_{cd}G^\beta_{de}
               V^{(1)}_{ef}G^\beta_{fb}
              +\cdots
\end{eqnarray}
Expanding the logarithm and dropping one term independent of the
perturbation $h^\pm_{\mu\nu}(x)$, the CTP effective action may be
perturbatively written as,
\begin{eqnarray}
   \Gamma^\beta_{CTP}[h^\pm_{\mu\nu}]
        & \ = \ &  S^{div}_g[h^+_{\mu\nu}] - S^{div}_g[h^-_{\mu\nu}]
                \nonumber \\
        &       & +{i\over2}Tr[ V^{(1)}_{+}G^\beta_{++}
                               -V^{(1)}_{-}G^\beta_{--}
                               +V^{(2)}_{+}G^\beta_{++}
                               -V^{(2)}_{-}G^\beta_{--}
                              ]
                \nonumber \\
        &       & -{i\over4}Tr[  V^{(1)}_{+}G^\beta_{++}
                                 V^{(1)}_{+}G^\beta_{++}
                               + V^{(1)}_{-}G^\beta_{--}
                                 V^{(1)}_{-}G^\beta_{--}
                               -2V^{(1)}_{+}G^\beta_{+-}
                                 V^{(1)}_{-}G^\beta_{-+}
                              ].
   \label{eq:effective action}
\end{eqnarray}
However, unlike the case of  \cite{CamHu} where $h_{\mu\nu}$
represented a perturbation about flat space and hence one had
knowledge of  exact  ``unperturbed" thermal propagators, in this
case, since the perturbation is about Schwarzschild spacetime,
exact expressions for the corresponding unperturbed propagators
$G^\beta_{ab}[h^\pm_{\mu\nu}]$ are not known. Therefore apart
from the approximation of computing the CTP effective action to
certain order in perturbation theory, an appropriate
approximation scheme for the unperturbed Green's functions is
also required. This feature manifested itself in York's
calculation of backreaction as well, where in writing the $<T_{\mu
\nu}>$ on the right hand side of the semiclassical Einstein
equation in the unperturbed Schwarzschild metric, he had to use
an approximate expression for $<T_{\mu\nu}>$ in the Schwarzschild
metric given by Page \cite{Page82}. The additional complication
here is that while to obtain $<T_{\mu\nu}>$ as in York's
calculation, the knowledge of only the thermal Feynman Green's
function is required, to calculate the CTP effective action one
needs the knowledge of the full matrix propagator, which involves
the Feynman, Schwinger and Wightman functions.

It is indeed possible to construct the full thermal matrix
propagator $G^\beta_{ab}[h^\pm_{\mu\nu}]$ based on Page's
approximate Feynman Green's function by using identities relating
the Feynman Green's function with the other Green's functions
with different boundary conditions. One can then proceed to
explicitly compute a CTP effective action and  the influence
functional based on this approximation. However, we desist from
delving into such a calculation for the following reason. Our
main interest in performing such a calculation is to identify and
analyze the noise term which is the new ingredient in the
backreaction. We have mentioned that the noise term gives a
stochastic contribution $T_{\mu\nu}^{\rm s}$ to the Einstein
Langevin equation (\ref{eq:effective stress tensor}). We had also
stated that this term is related to the variance of fluctuations
in $T_{\mu\nu}$, i.e, schematically, to $< T^2_{\mu\nu}>$.
However, a calculation of $< T^2_{\mu\nu}>$ in the Hartle-Hawking
state in a Schwarzschild background using the Page approximation
was performed by Phillips and Hu \cite{PH1,PH2,PH3} and it was
shown that though the approximation is excellent as far as $<
T_{\mu\nu}>$ is concerned, it gives unacceptably large errors for
$< T^2_{\mu\nu}>$ at the horizon. In fact, similar errors will be
propagated in the non-local dissipation term as well, because
both terms originate from the same source, that is, they come
from the last trace term in (\ref{eq:effective action}) which
contains terms quadratic in the Green's function. However, the
Influence Functional or CTP formalism itself does not depend on
the nature of the approximation, so we will attempt to exhibit
the general structure of the calculation without resorting to a
specific form for the Greens function and conjecture on what is
to be expected. A more accurate computation can be performed
using this formal structure once a better approximation becomes
available.

If we denote the difference and the sum of the perturbations
$h^\pm_{\mu\nu}$  by $[h_{\mu\nu}] \equiv h^+_{\mu\nu} -
h^-_{\mu\nu}$ and $\{h_{\mu\nu}\} \equiv h^+_{\mu\nu} +
h^-_{\mu\nu}$, respectively, the influence functional form of the
thermal CTP effective action may be written to second order in
$h_{\mu\nu}$ as \cite{CamHu},
\begin{eqnarray}
   \Gamma^\beta_{CTP}[h^\pm_{\mu\nu}]
        & \ \simeq \ & {1\over2\ell^2_P}\int d^4x\ d^4x'\
                       [h_{\mu\nu}](x){L}_{(o)}^{\mu\nu,\alpha\beta}(x,x')
                       \{h_{\alpha\beta}\}(x')
                     \nonumber \\
        &            &+{1\over2}\int d^4x\
                       [h_{\mu\nu}](x)T^{\mu\nu}_{(\beta)}
                     \nonumber \\
        &            &+{1\over2}\int d^4x\ d^4x'\
                       [h_{\mu\nu}](x){H}^{\mu\nu,\alpha\beta}(x,x')
                       \{h_{\alpha\beta}\}(x')
                     \nonumber \\
        &            &-{1\over2}\int d^4x\ d^4x'\
                       [h_{\mu\nu}](x)D^{\mu\nu,\alpha\beta}(x,x')
                       \{h_{\alpha\beta}\}(x')
                     \nonumber \\
        &            &+{i\over2}\int d^4x\ d^4x'\
                       [h_{\mu\nu}](x)N^{\mu\nu,\alpha\beta}(x,x')
                      [h_{\alpha\beta}](x').
\label{CTPbh}
\end{eqnarray}

The first line is the Einstein-Hilbert action to second order in
the perturbation $h^\pm_{\mu\nu}(x)$ and $\kl{L}{(o)}{x}$ is a
symmetric kernel, {\sl i.e.} $\kl{L}{(o)}{x,x'}$ =
$\kl{L}{(o)}{x',x}$. The second is a local term linear in
$h^\pm_{\mu\nu}(x)$. $T_{(\beta)}^{\mu\nu}$ represents the zeroth
order contribution to $<T_{\mu\nu}>$ and far away from the hole
it takes the form of the stress tensor of massless scalar
particles at temperature $\beta^{-1}$. The third and fourth terms
constitute the remaining quadratic component of the real part of
the effective action. The kernels $H^{\mu\nu,\alpha\beta}(x,x')$
and $D^{\mu\nu,\alpha\beta}(x,x')$ are respectively even and odd
in $x,x'$. The last term gives the imaginary part of the
effective action and the kernel $N(x,x')$ is symmetric. This is
the general structure of the CTP effective action arising from
the calculation of the traces in equation (\ref{eq:effective
action}). Of course, to write down explicit expressions for the
non-local kernels one requires the input of the explicit form of
$G^\beta_{ab}[h^\pm_{\mu\nu}]$ , which we have not used. In spite
of this limitation we can make some interesting observations from
this effective action. Connecting this thermal CTP effective
action to the influence functional via equation (\ref{CTPIFconn})
we see that the nonlocal imaginary term containing the kernel
$N^{\mu\nu,\alpha\beta}(x,x')$ is responsible for the generation
of the stochastic noise term in the Einstein-Langevin equation
and the real non-local term containing kernel
$D^{\mu\nu,\alpha\beta}(x,x')$ is responsible for the non-local
dissipation term. The Einstein-Langevin equation can be generated
from equation (\ref{CTPstoch}) by first constructing the improved
semiclassical effective action in accordance with (\ref{Seffxi})
and deriving the equation of motion (\ref{stochem}) by taking a
functional derivative of the above effective action with respect
to $[h_{\mu\nu}]$ and equating it to zero. With the
identification of noise and dissipation kernels, one can use them
to write down a Fluctuation-Dissipation relation (FDR) analogous
to (\ref{FDR}) in the context of black holes.

\subsection{Einstein-Langevin equation}

In this section we show how  a semiclassical
Einstein-Langevin equation  can be derived from the previous
thermal CTP effective action. This equation depicts the stochastic
evolution of the perturbations of the black hole under the influence of the
fluctuations of the thermal scalar field.

The influence functional
${\cal F} \equiv \exp (iS_{IF})$
previously introduced in section {\ref{sec:open}} can be
written in terms of the the CTP effective action $\Gamma^\beta_{CTP}[h^\pm_{\mu\nu}]$ derived in equation (\ref{CTPbh}) as follows using the connection given by equation (\ref{CTPIFconn})

\begin{equation}
   {\cal F}
        \ = \ \exp i\left( Re \{ \Gamma^\beta_{CTP}[h^\pm_{\mu\nu}] \}
                          +{i\over2}\int d^4x\ d^4x'\
                               [h_{\mu\nu}](x)\kl{N}{}{x-x'}
                               [h_{\alpha\beta}](x')
                    \right),
\end{equation}
where $Re\{\ \}$ denotes taking the real part. Following
\cite{if,ssg}, we can interpret the real part of
the influence functional as the characteristic functional of a
non-dynamical stochastic variable $j^{\mu\nu}(x)$,
\begin{equation}
   \Phi([h_{\mu\nu}])
        \ = \ \exp \left( -{1\over2}\int d^4x\ d^4x'\
                           [h_{\mu\nu}](x)\kl{N}{}{x-x'}
                           [h_{\alpha\beta}](x')
                   \right).
   \label{eq:cf}
\end{equation}
This classical stochastic field represents probabilistically the quantum
fluctuations of the matter field and is responsible for the
dissipation of the gravitational field. By definition, the
above characteristic functional is the functional Fourier transform of
the probability distribution functional ${\cal P}[j^{\mu\nu}]$ with
respect to $j^{\mu\nu}$,
\begin{equation}
   \Phi([h_{\mu\nu}])
        \ = \ \int {\cal D}j^{\mu\nu}\ {\cal P}[j^{\mu\nu}]\
              e^{i\int d^4x\ [h_{\mu\nu}](x)j^{\mu\nu}(x) }.
   \label{eq:cf_pdf}
\end{equation}
Using (\ref{eq:cf}) one can easily see that the probability
distribution functional is related with the noise kernel by the formal
expression,
\begin{equation}
   {\cal P}[j^{\mu\nu}]
        \ = \ {\exp \left( -{1\over2}\int d^4x\ d^4x'\
                           j_{\mu\nu}(x)[\kl{N}{}{x-x'}]^{-1}
                           j_{\alpha\beta}(x')
                   \right)
               \over
               \int {\cal D}j^{\mu\nu}\
               \exp \left( -{1\over2}\int d^4x\ d^4x'\
                           j_{\mu\nu}(x)[\kl{N}{}{x-x'}]^{-1}
                           j_{\alpha\beta}(x')
                   \right)
              }.
   \label{eq:pdf}
\end{equation}
For an arbitrary functional of the stochastic field ${\cal
A}[j^{\mu\nu}]$, the average value with respect to the previous
probability distribution functional is defined as the functional
integral
 \be
 \langle {\cal A}[j^{\mu\nu}] \rangle_j
 \equiv \int {\cal D}[j^{\mu\nu}]\
             {\cal P}[j^{\mu\nu}]
             {\cal A}[j^{\mu\nu}].
 \te
In terms of this stochastic average the influence functional can be
written as
${\cal F} =
 \langle \exp\left(i\Gamma^{st}_{CTP}[h^\pm_{\mu\nu}]
             \right)
 \rangle_j$, where $\Gamma^{st}_{CTP}[h^\pm_{\mu\nu}]$ is the
stochastic effective action
\begin{equation}
   \Gamma^{st}_{CTP}[h^\pm_{\mu\nu}]
        \ \equiv \  Re \{ \Gamma^\beta_{CTP}[h^\pm_{\mu\nu}] \}
                   +\int d^4x\ [h_{\mu\nu}](x) j^{\mu\nu}(x).
   \label{eq:mea}
\end{equation}
Clearly, because of the quadratic definition of the characteristic
functional (\ref{eq:cf}) and its relation with the probability
distribution functional (\ref{eq:cf_pdf}), the field $j^{\mu\nu}(x)$ is a
zero mean Gaussian stochastic variable. This means that its two-point
correlation function, which is given in terms of the noise kernel by
\begin{equation}
   \langle j^{\mu\nu}(x) j^{\alpha\beta}(x') \rangle_j
        \ = \ \kl{N}{}{x-x'},
   \label{eq:correlation}
\end{equation}
completely characterizes the stochastic process. The Einstein-Langevin
equation follows from taking the functional derivative of the
stochastic effective action (\ref{eq:mea}) with respect to
$[h_{\mu\nu}](x)$ and  imposing $[h_{\mu\nu}](x) = 0$
 \cite{ssg} as an analog of (\ref{stochem}). In our case, this leads to
\begin{equation}
   {1\over\ell^2_P}
   \int d^4x'\ \kl{L}{(o)}{x-x'} h_{\alpha\beta}(x')
  +{1\over2}\ T^{\mu\nu}_{(\beta)}
  +\int d^4x'\ \left( \kl{H}{}{x-x'}
                     -\kl{D}{}{x-x'}
               \right) h_{\alpha\beta}(x')
  +j^{\mu\nu}(x)
        \ = \ 0.
\end{equation}

In the far field limit this equation should reduce to that
obtained by Campos and Hu \cite{CamHu}: For gravitational
perturbations $h^{\mu\nu}$ defined in (\ref{eq:def bar h}) under
the harmonic gauge $\bar h^{\mu\nu}_{\,\,\,\,\, ,\nu} = 0$, their
Einstein-Langevin equation is given by
\begin{equation}
   \Box\bar h^{\mu\nu}(x)
         + \ell^2_P
               \left\{ T^{\mu\nu}_{(\beta)}
                      +2P_{\rho\sigma,\alpha\beta}
                       \int d^4x'\ \left( \kl{H}{}{x-x'}
                                         -\kl{D}{}{x-x'}
                                   \right)\bar h^{\rho\sigma}(x')
                      +2j^{\mu\nu}(x)
               \right\} = 0,
\end{equation}
where the tensor $P_{\rho\sigma,\alpha\beta}$ is given by
\begin{equation}
   P_{\rho\sigma,\alpha\beta}
        \ = \ {1\over2}\left( \eta_{\rho\alpha}\eta_{\sigma\beta}
                             +\eta_{\rho\beta}\eta_{\sigma\alpha}
                             -\eta_{\rho\sigma}\eta_{\alpha\beta}
                       \right).
\end{equation}
The expression for  $P_{\rho\sigma,\alpha\beta}$ in the near
horizon limit of course cannot be expressed in such a simple form.
Note that this differential stochastic equation includes a
non-local term responsible for the dissipation of the
gravitational field and a noise source term which accounts for
the fluctuations of the quantum field . Note also that this
equation in combination with the correlation for the stochastic
variable (\ref{eq:correlation}) determine the two-point
correlation for the stochastic metric fluctuations $\langle \bar
h_{\mu\nu}(x) \bar h_{\alpha\beta}(x') \rangle_j$
self-consistently.


\section{CTP Effective Action in terms of Bogolyubov Coefficients}

Before we draw further implications from this structural form of
the CTP effective action and the stochastic effective action on
the effect of fluctuations and dissipation, we would like to add
another perspective to this problem by considering a different
aspect of the interpretation of the stochastic fluctuation term,
namely its connection with the fluctuation in the number of
particles produced. Studies of the Influence Functional in the
context of semiclassical cosmology
\cite{calmaz,pazsinha,noisefluct,HuSin} show that the Influence
Functional can be written in terms of the Bogolyubov coefficients
that represent particle production in the specific cosmological
background. The noise kernel that arises in this Influence
functional can be related to the difference between the number of
particles produced along alternative histories. The noise term in
the Einstein-Langevin equation can in turn be related to the
fluctuations in the number of particles produced by the
cosmological background. The relationship between noise and
particle number fluctuations has also been emphasized in
\cite{Kang}. Gaining insight from this approach, we have
attempted to cast the Influence Functional or equivalently the CTP
effective action for the black hole in a similar form in terms of
Bogolyubov coeffecients. In this case these coefficients refer to
the production of particles by the black hole background. In this
section we will describe such a derivation.

The CTP effective action, which is a functional of two metric
histories, $g_{\mu\nu}^{(+)}$ and $g_{\mu \nu}^{(-)}$, may be
expressed in terms of the Bogolyubov coefficients that relate the
``in'' vacuum state to the ``out'' vacuum state in these two
histories. Such a representation has been derived earlier
\cite{noisefluct,HuMataczParam} for cosmological metrics. For
these cases, the Bogolyubov coefficient matrices (denoted by
$\alpha$ and $\beta$) are diagonal. In the general case (and
specifically for black hole metrics), the Bogolyubov coefficient
matrices are not diagonal, and the derivation of the CTP
effective action is a bit more involved.

Here, we follow the vacuum-to-many-particle-amplitude approach of
DeWitt  \cite{dw75}, who used this method to derive an expression
for the ``in-out'' effective action in terms of Bogolyubov
coefficients.

The CTP generating functional is the product of the time-evolution
amplitude
from a vacuum state in the distant past (``in'' vacuum) to an arbitrary
state
in the distant future in the metric $g_{\mu \nu}^{(+)}$ with the
time-reversed
amplitude from the same arbitrary state in the distant future to the
``in''
vacuum state in the metric $g_{\mu \nu}^{(-)}$, summed over all
arbitrary states
in the distant future. We thus obtain
\begin{equation}
e^{i\Gamma_{\rm CTP}}=\sum_{n,i_1,...i_n}\frac{1}{n!}\,
\hphantom{}_{\rm in}\langle 0 \mid 1_{i_1},1_{i_2},...1_{i_n}
\rangle_{\rm out}^{(+)}\,\,\hphantom{}_{\rm out}\langle
1_{i_1},1_{i_2},..1_{i_n}
\mid 0\rangle_{\rm in}^{(-)},
\label{rav1}
\end{equation}
where $\mid 1_{i_1},1_{i_2},...1_{i_n}\rangle_{\rm out}$ is the state
containing
one particle each in modes $i_1, i_2,\ldots,i_n$ in the ``out''
particle
model,
the $(+)$ and $(-)$ superscripts refer to the two metric histories, and
the $n!$
in the denominator avoids overcounting of states.

DeWitt  \cite{dw75} writes down the amplitude to evolve from the
``in'' vacuum state to the state containing single particles in
$m$ modes in the distant future, in terms of the Bogolyubov
coefficients, as follows:
\begin{equation}
\hphantom{}_{\rm out}\langle 1_{i_1},1_{i_2},..1_{i_m} \mid
0\rangle_{\rm in}^{(-)}
=e^{i\Gamma_{\rm in-out}^{(-)}}\sum_{n=0}^{\infty}\frac{i^{n/2}}{n!}
\sum_{j_1,j_2,...,j_n}V_{j_1,...,j_n}^{(-)}\hphantom{}_{\rm out}
\langle 1_{i_1},1_{i_2},...,1_{i_m}\mid
1_{j_1},1_{j_2},...,1_{j_n}\rangle_{\rm out},
\label{rav2}
\end{equation}
where
\begin{eqnarray}
V_{i_1,...,i_n}&=&\sum_p V_{i_1 i_2}\cdots V_{i_{n-1}i_n}, ~~~ n
\,\,{\rm even} \label{rav3}\\
&=& 0, ~~~ n \,\,{\rm odd}
\end{eqnarray}
with
\begin{equation}
V_{ij}=i\sum_k \beta_{ki}^{\ast}\alpha_{jk}^{-1},
\label{rav4}
\end{equation}
and the sum in Eq. (\ref{rav3}) is over all $n!/(2^{n/2}(n/2)!)$
distinct pairings
of the labels $i_1,...,i_n$.

Note that
\begin{equation}
\hphantom{}_{\rm out}\langle 1_{i_1},1_{i_2},...,1_{i_m}\mid
1_{j_1},1_{j_2},...,1_{j_n}
\rangle_{\rm out} = \delta_{mn} \sum_{\gamma} \delta_{i_1
j_{\gamma(1)}}\ldots
\delta_{i_n j_{\gamma(n)}}
\label{rav5}
\end{equation}
where the sum is now over all possible permutations $\gamma$ of the
indices $1,\ldots,n$.
This allows us to simplify Eq. (\ref{rav2}) to yield
\begin{eqnarray}
\hphantom{}_{\rm out}\langle 1_{i_1},1_{i_2}\ldots,1_{i_m} \mid
0\rangle_{\rm in}^{(-)}
&=&e^{i\Gamma_{\rm in-out}^{(-)}}
\frac{i^{m/2}}{m!}\sum_{\gamma}V_{i_{\gamma(1)}
\ldots i_{\gamma(m)}} \nonumber \\
&=& e^{i\Gamma_{\rm in-out}^{(-)}} i^{m/2}V_{i_1\ldots i_m},
\label{rav6}
\end{eqnarray}
where the second equality follows from the first one because each
distinct pairing of
indices has been repeated $m!$ times in the sum over all permutations.

Equation (\ref{rav6}) and its complex conjugate in the $(+)$ history
may
be substituted
in Eq. (\ref{rav1}) to yield
\begin{eqnarray}
e^{i\Gamma_{\rm CTP}} &=& e^{i\Gamma_{\rm
in-out}^{(-)}}e^{-i\Gamma_{\rm
in-out}^{(+)\,\ast}}
\sum_{n=0}^{\infty} \frac{1}{n!}\sum_{i_1,\ldots,i_n}
V_{i_1,\ldots,i_n}^{(-)}V_{i_1, \ldots,i_n}^{(+)\,\ast} \nonumber \\
&=& ({\rm det}\,\alpha^{(-)})^{-1/2}({\rm
det}\,\alpha^{(+)\ast})^{-1/2}\sum_{n=0}^{\infty}
\frac{1}{n!}\sum_{i_1,\ldots,i_n}V_{i_1,\ldots,i_n}^{(-)}V_{i_1,
\ldots,i_n}^{(+)\,\ast},
\label{rav7}
\end{eqnarray}
where the representation of the in-out effective action in terms
of the $\alpha$ Bogolyubov coefficient matrix has been derived in
\cite{dw75}. In order to evaluate the infinite sum in the above
equations, we may expand the $ V_{i_1,\ldots,i_n}^{(-)}V_{i_1,
\ldots,i_n}^{(+)\,\ast}$ term using Eq. (\ref{rav3}) and collect
like powers of $V^{(-)}V^{(+)\dag}$. We then observe that the
general term in the infinite sum is of the form
\begin{equation}
\sum_{r=1}^n \frac{1}{2^r r!}\sum_{\left\{m_i \geq 1;\sum_{i=1}^r m_i
=n\right\}}\frac{{\rm Tr}(V^{(-)}V^{(+)\dag})^{m_1}\cdots{\rm
Tr}(V^{(-)}V^{(+)\dag})^{m_r}}
{\prod_{i=1}^rm_i},
\label{rav8}
\end{equation}
where the second sum is over all possible values of $r$ natural numbers
$m_i, i=1,\ldots,r$ whose sum is $n$.

One may readily verify that (\ref{rav8}) is also the general term in
the
expansion of
\begin{equation}
{\rm det}(1-V^{(-)}V^{(+)\dag})^{-1/2}=
{\rm exp}\left(-{1 \over 2}{\rm Tr}\,{\rm
ln}(1-V^{(-)}V^{(+)\dag})\right).
\label{rav9}
\end{equation}
We thus obtain the CTP effective action in terms of Bogolyubov
coefficients, as
\begin{eqnarray}
e^{i\Gamma_{\rm CTP}}&=&({\rm det}\,\alpha^{(-)})^{-1/2}({\rm det}\,
\alpha^{(+)\ast})^{-1/2}{\rm det}(1-V^{(-)}V^{(+)\dag})^{-1/2}\nonumber
\\
&=& {\rm
det}\left(\alpha^{(+)\ast}\alpha^{(-)}-\beta^{(-)\ast}\beta^{(+)}\right)^{-1/2},
\label{rav10}
\end{eqnarray}
where the second equality follows from Eq. (\ref{rav4}) and the use of
${\rm det}(AB)={\rm det}(BA)$.

Note that the above effective action is the full unrenormalized
CTP effective action and therefore the determinant is divergent.
In practice, these divergences must be isolated and absorbed into
the metric and the curvatures in the standard manner.

Our formula (4.11) reduces in a straightforward way to the
formula for the cosmological case given in [59]. Note that in
this section we have assumed the ``in'' vacuum state to be a pure
state. Therefore, (4.11) does not directly apply to situations
where the ``in'' state is a density matrix, as in, for example,
the Hartle-Hawking vacuum state discussed in the previous
section. Generalizations of (4.11) to arbitrary initial states
(i.e., pure or mixed) follow a similar line of reasoning as
discussed here and will be given in a future work. The restricted
version of the CTP effective action given here can, however, be
used to study evaporating black holes where the initial state is
the Unruh vacuum, and to derive an Einstein-Langevin equation for
the evaporating black hole.

The Bogolyubov coefficient representation may also serve as a
seed for approximation schemes that are based on the Bogolyubov
coefficients rather than on the Green's functions. Viewing the
noise or fluctuation contribution in terms of fluctuations in
particle number also offers the interesting possibility of
interpretation in terms of the \textbf{isothermal compressibility
of the vacuum} by exploiting the thermodynamic properties of
black holes.

\newpage
\section{Discussions}

At this point it is a good idea to take stock of what has been
achieved by taking this strategy and also compare the results
with earlier approaches. By setting up the CTP effective
action/Influence Functional for this model system, we have a
formal causal framework for dealing with the backreaction of
quantum fields on a black hole within the limitations of
linearization and the special symmetries of the model. At
present,though we know that with the Page approximation some
backreaction terms in the Einstein-Langevin equation become
unreliable, as and when better approximations to the
Schwarzschild Green's functions become available this Influence
Functional can be used to yield an Einstein-Langevin equation for
black hole back reaction which encompasses the effects of
dissipation as well as noise and fluctuations.

If we confine ourselves to Page's approximation and derive the
equation of motion using  (\ref{CTPem}) rather than
({\ref{CTPstoch}), i.e, without the stochastic term, we expect to
recover York's semiclassical Einstein's equation if one retains
only the zeroth order contribution, i.e, the first two terms in
the expression for the CTP effective action in equation
(\ref{CTPbh}) Thus, this offers a new route to arrive at York's
semiclassical Einstein's equations. Not only is it a derivation
of York's result from a different point of view, but it also
shows how his result arises as an appropriate limit of a more
complete framework, i.e, it arises when one averages over the
noise. Another point worth noting is that our treatment  will
also yield a non-local dissipation term arising from the fourth
term in equation (\ref{CTPbh}) in the CTP effective action which
is absent in York's treatment. This difference arises primarily
due to the difference in the way backreaction is treated at the
level of iterative approximations on the equation of motion
versus the treatment in the effective action framework. In York's
treatment, the Einstein tensor is computed to first order in
perturbation theory , while $<T_{\mu\nu}>$ on the right hand side
of the semiclassical Einstein equation is replaced by the zeroth
order term. In the effective action treatment the full effective
action is computed to second order in perturbation, and hence
includes the higher order non-local terms.

The other important conceptual point that comes to light from this
approach is that related to the Fluctuation-Dissipation Relation.
Here it is clearly seen that the backreaction of quantum fields
on black holes consists of two forms -- dissipation and
fluctuation or noise, which contribute respectively to the real
and imaginary parts of the Influence functional as embodied in
the dissipation and noise kernels. Again drawing the analogy with
Brownian motion, these correspond to the dissipation of the
energy of the Brownian particle as it approaches equilibrium, and
the fluctuations at equilibrium. These are connected by the
Fluctuation - Dissipation relation given by (\ref{FDR}). Pursuing
the analogy we conjecture that a FDR similar to (\ref{FDR}) will
also exist between the noise and dissipation kernels for the
black hole case that we have described. This reveals an
interesting connection between black holes interacting with
quantum fields and non-equilibrium statistical mechanics. The
existence of a FDR for the black hole case has been discussed by
some authors previously  \cite{CanSci,Mottola}. In  \cite{Vishu}
we have described in some detail how our approach differs from
those of previous authors. We refer the reader to  \cite{Vishu}
for details.

There are, of course, limitations of our program. One obvious one
is that we have to confine ourselves to small perturbations about
the static Schwarzschild background. As a result we cannot hope
to address questions about the fully dynamical collapse problem.
However, it will allow us to study the stability of the black
hole under the influence of stochastic fluctuations of the energy
momentum tensor dictated by the noise terms.

Another limitation we have already discussed before is the
problem of finding a reliable approximation to the Schwarzschild
thermal Green's function to explicitly compute the noise and
dissipation kernels. This limits our ability to present explicit
analytical expressions for these kernels. One possibility is to
try to work on improving on Page's approximation by retaining
terms to higher order. A less ambitious first step could be to
confine attention to the horizon  and using approximations that
are restricted to near the horizon and work out the Influence
Functional in this regime. This is currently being pursued by us.
The Influence Functional in the far field regime has already been
worked out in  \cite{CamHu}.

The other shortcoming is the following. Though we have allowed for
backreaction effects to modify the initial state in the sense
that  the temperature of the Hartle-Hawking state gets affected
by the backreaction, we have essentially confined our analysis to
a Hartle-Hawking thermal state of the field. This analysis does
not directly extend to a more general class of states, for
example to the case where the initial state of the field is in
the Unruh vacuum. Thus we will not be able to comment on issues
of the stability of an isolated radiating black hole under the
influence of stochastic fluctuations.

It is also pertinent to mention the connection of our work with
that of some other authors who have also pursued the viewpoint
that the black hole metric is a stochastically fluctuating
quantity and have studied its effect on Hawking radiation. For
example, Casher et al  \cite{CEIMP} and Sorkin \cite{Sorkin} have
concentrated on the issue of fluctuations of the horizon induced
by a fluctuating metric. Casher et al \cite{CEIMP} considers the
fluctuations of the horizon induced by the ``atmosphere" of high
angular momentum particles near the horizon, while Sorkin
\cite{Sorkin} calculates fluctuations of the shape of the horizon
induced by the quantum field fluctuations under a Newtonian
approximations. Both group of authors come to the conclusion that
horizon fluctuations become large at scales much larger than the
Planck scale (note Ford and Svaiter \cite{Ford} later presented
results contrary to this claim). However, though these works do
deal with backreaction, the fluctuations considered do not arise
as an explicit stochastic noise term as in our treatment. It may
be worthwhile exploring the horizon fluctuations induced by the
stochastic metric in our model and comparing the conclusions with
the above authors. Barrabes et al \cite{Barrabes} (also see
\cite{Parentani} for work in similar vein) have considered the
propagation of null rays and massless fields in a black hole
fluctuating geometry and have shown that the stochastic nature of
the metric leads to a modified dispersion relation and helps to
confront the trans-Planckian frequency problem. However, in this
case the stochastic noise is put in by hand and does not
naturally arise from coarse graining as in the quantum open
systems approach.  It also does not take backreaction into
account. It will be interesting to explore how a stochastic black
hole metric, arising as a solution to the Einstein-Langevin
equation, hence fully incorporating backreaction would affect the
trans-Planckian problem.

Ford and his collaborators  \cite{Ford} have also explored the
issue of metric fluctuations in detail and in particular have
studied the fluctuations of the black hole horizon induced by
metric fluctuations. However, the fluctuations they considered
are in the context of a fixed background and do not relate to the
backreaction.

Another work originating from the same vein of stochastic gravity
but not complying with the backreaction spirit is that of Hu and
Shiokawa \cite{HuShio}, who study novel effects associated with
electromagnetic wave propagation in a Robertson-Walker universe
and the Schwarzschild spacetime with a small amount of given
metric stochasticity. For the Schwarzschild metric, they find
that time-independent randomness can decrease the total
luminosity of Hawking radiation due to multiple scattering of
waves outside the black hole and gives rise to event horizon
fluctuations and fluctuations in the Hawking temperature. The
stochasticity in a background metric in their work is assumed
rather than derived (from quantum field fluctuations, as in this
work) and so is not in the same spirit of backreaction. But it is
interesting to compare their results with that of backreaction,
so one can begin to get a sense of the different sources of
stochasticity and their weights (see, e.g., \cite{stogra} for a
list of possible sources of stochasticity.)

In a subsequent paper Shiokawa \cite{Shio} showed that the scalar
and spinor waves in a stochastic spacetime behave similarly to
the electrons in a disordered system. Viewing this as a quantum
transport problem, he expressed the conductance and its
fluctuations in terms of a nonlinear sigma model in the closed
time path formalism and showed that the conductance fluctuations
are universal, independent of the volume of the stochastic region
and the amount of stochasticity. This result can have significant
importance in characterizing the mesoscopic behavior of
spacetimes resting between the semiclassical and the quantum
regimes.

The stochastic approach to the study of black hole backreaction
thus has a very rich structure and opens up many new avenues of
inquiry. In particular it provides the proper platform and
framework to launch a new program of research into the
\textbf{nonequilibrium black hole thermodyamics}.\\

\noindent {\bf Acknowledgements} BLH wishes to thank Toni
Campos,  Nicholas Phillips and K. Shiokawa for collaborations on
topics related to black hole backreaction and fluctuations, and
Drs. Paul Anderson, Larry Ford and Enric Verdaguer for general
discussions on related problems. This work is supported in part
by NSF grant PHY98-00967.

\end{document}